\documentclass[superscriptaddress, aps, prx, twocolumn, longbibliography]{revtex4-2}

\usepackage{graphicx}
\usepackage{physics}
\usepackage{amsmath}
\usepackage{array}
\usepackage{siunitx}
\usepackage{hyperref}

\usepackage{natbib}

\usepackage{gensymb}
\usepackage{bm}
\usepackage{amssymb}

\newcommand{\Cq}{\mathcal{C}_q}
\newcommand{\gs}{\gamma_\text{s}}
\newcommand{\Gb}{\Gamma_\text{lib}}
\newcommand{\nth}{n_\text{th}}
\newcommand{\gb}{\gamma_\text{lib}}
\newcommand{\Ob}{\omega_\text{lib}}
\newcommand{\kB}{k_\text{B}}

\usepackage{orcidlink}

\begin{document}

\title{On-chip levitation of ferromagnetic microparticles}

\date{April 2026}

\author{Martijn Janse\,\orcidlink{0000-0002-3877-9019}}
\author{M. Luisa Mattana\,\orcidlink{0009-0003-1793-0378}}
\author{Julian van Doorn\,\orcidlink{0009-0004-5892-8001}}
\author{Eli van der Bent\,\orcidlink{0009-0001-4270-8221}}
\author{Richard Wagner\,\orcidlink{0009-0009-8251-4462}}
\author{Robert Smit\,\orcidlink{0000-0003-2576-4851}}
\author{Bas Hensen\,\orcidlink{0000-0003-1754-6029}} \email{hensen@physics.leidenuniv.nl}
\address{Leiden Institute of Physics, Leiden University, P.O. Box 9504, 2300 RA Leiden, The Netherlands}

\begin{abstract}
Levitation of microscopic objects in vacuum combines exceptional environmental isolation with precise control of their dynamics, pushing the limits of sensing and macroscopic quantum physics. In particular, magnetic levitation allows a large range of particle sizes, while avoiding detrimental effects from high-intensity optical trapping beams and electric field noise. However, existing diamagnetic and Meissner levitation approaches are typically constrained by low mechanical eigenfrequencies, limited integrability with other systems due to bulky coils or magnets, and, for Meissner levitation, the need for cryogenic operation. Here, we demonstrate a room-temperature on-chip magnetic levitation platform capable of stably levitating a nanogram (6.5 micrometer radius) ferromagnetic microsphere. The platform is scalable and tunable, and supports librational modes with eigenfrequencies exceeding 10 kHz. Further miniaturization and coupling to solid-state spin qubits could enable cooling to the quantum ground state. Beyond quantum experiments, this architecture enables integrated precision sensing and studies of isolated ferromagnet thermodynamics.
\end{abstract}

\maketitle

Levitation in vacuum is increasingly recognized as a versatile platform for advancing both fundamental science and technological applications \cite{gonzalez-ballestero_levitodynamics_2021}. Recent advances have enabled precise control of isolated systems, allowing accurate measurement and manipulation of translational and rotational degrees of freedom \cite{millen_optomechanics_2020, dania_ultrahigh_2024, schmidt_remote_2024, delord_spin-cooling_2020, van_der_laan_sub-kelvin_2021, zielinska_controlling_2023}. These capabilities open access to previously unexplored regimes, including ultra‑sensitive force and acceleration sensing \cite{wang_mechanical_2024, ranjit_zeptonewton_2016, hebestreit_sensing_2018}, searches for dark matter candidates \cite{amaral_first_2025, ahrens_levitated_2025, higgins_maglev_2024, carney_mechanical_2021, kalia_ultralight_2024}, and the generation of macroscopic quantum states \cite{romero-isart_large_2011, scala_matter-wave_2013, wan_free_2016, margalit_realization_2021, bose_massive_2025}. Although levitation can be achieved using optical, electrical, or magnetic trapping potentials, most progress has focused on optical levitation. Recent breakthroughs include feedback and sideband cooling of translational \cite{delic_cooling_2020, tebbenjohanns_quantum_2021, magrini_real-time_2021, piotrowski_simultaneous_2023} and librational \cite{dania_high-purity_2025, troyer_quantum_2026} modes to the quantum ground state, and state expansion \cite{kamba_revealing_2023, bonvin_state_2024, rossi_quantum_2025, steiner_free_2025,kamba_quantum_2025,tomassi_accelerated_2026, mattana_trap--trap_2026}.\\

Exploiting levitation to study the interplay between quantum mechanics and gravity requires confining increasingly massive objects \cite{bose_massive_2025, bose_spin_2017, bassi_models_2013, oppenheim_postquantum_2023, gonzalez-ballestero_levitodynamics_2021,fuchs_measuring_2024}. In optical levitation, gravitational forces must be counteracted by laser-induced optical force, which ultimately limits the levitated mass due to laser‑induced evaporation \cite{monteiro_optical_2017}. Electrical traps circumvent this limitation but introduce additional decoherence arising from charge noise \cite{brownnutt_ion-trap_2015}. Therefore, in order to levitate heavier neutral objects, magnetic levitation has gained more attention. Diamagnetic levitation at room temperature \cite{slezak_cooling_2018} and cryogenic temperatures \cite{gieseler_single-spin_2020, vinante_ultralow_2020, gutierrez_latorre_superconducting_2023, fuchs_measuring_2024, schmidt_remote_2024, timberlake_linear_2024} have been demonstrated, trapping micro- up to  milligram masses with high quality factors. So far, these approaches suffer from weak confining potentials resulting in modest eigenfrequencies on the order of 10-100 Hz. In addition, their physical implementation requires bulky coils or magnets, or the need for cryogenic operation.\\ 

Recently, a novel scheme for room-temperature magnetic levitation, known as the magnetic Paul trap, has been proposed \cite{cornell_multiply_1991, sackett_magnetic_1993, perdriat_planar_2023}. A proof-of-principle demonstration using alternating currents has been realized on a printed circuit board \cite{janse_characterization_2024}. This platform has a number of advantageous properties. First, because the magnetic dipole moment to mass ratio $\mu/m$ of the levitating ferromagnet is independent of particle size, the underlying trapping mechanism is scale invariant \cite{sackett_magnetic_1993}. As a result, the same trap can be used to levitate objects spanning macroscopic to nanoscopic scales, enabling systematic size-dependent studies under identical conditions, such as precision tests of short-range gravity~\cite{fuchs_measuring_2024, westphal_measurement_2021} and the quantum-to-classical transition~\cite{bose_massive_2025}. Second, the levitated particle is oriented along a well-defined axis, set by a static external magnetic field. This confinement of the rotational degree of freedom results in a high-frequency librational eigenmode, ideally suited for coupling to spin-based quantum systems \cite{gieseler_single-spin_2020, huillery_spin_2020}. Finally, the platform can be integrated on-chip, allowing nanoscale engineering, arrays of traps and straightforward integration with other quantum systems, including nitrogen-vacancy center spin defects in diamond and superconducting circuits.\vspace{4pt}

Here we present the experimental realization of such an on-chip magnetic Paul trap, stably levitating a nanogram (6.5 $\mu $m radius) ferromagnetic microsphere in vacuum with translational eigenmodes up to 500 Hz and librational eigenfrequencies exceeding 10 kHz. We demonstrate the tunability of the platform and study thermodynamics induced by the readout laser at low gas pressures. We find that the damping of both translational and librational modes is set by background‑gas interactions down to the lowest verifiable gas pressures in our setup at room temperature. Finally, we propose that with further miniaturization and integration with solid-state spin qubits, the librational eigenmodes can operate in the single-spin strong-coupling regime, allowing sideband cooling to the quantum ground state.

\begin{figure*}[hbt!]
    \centering
    \includegraphics[page=1,scale=1]{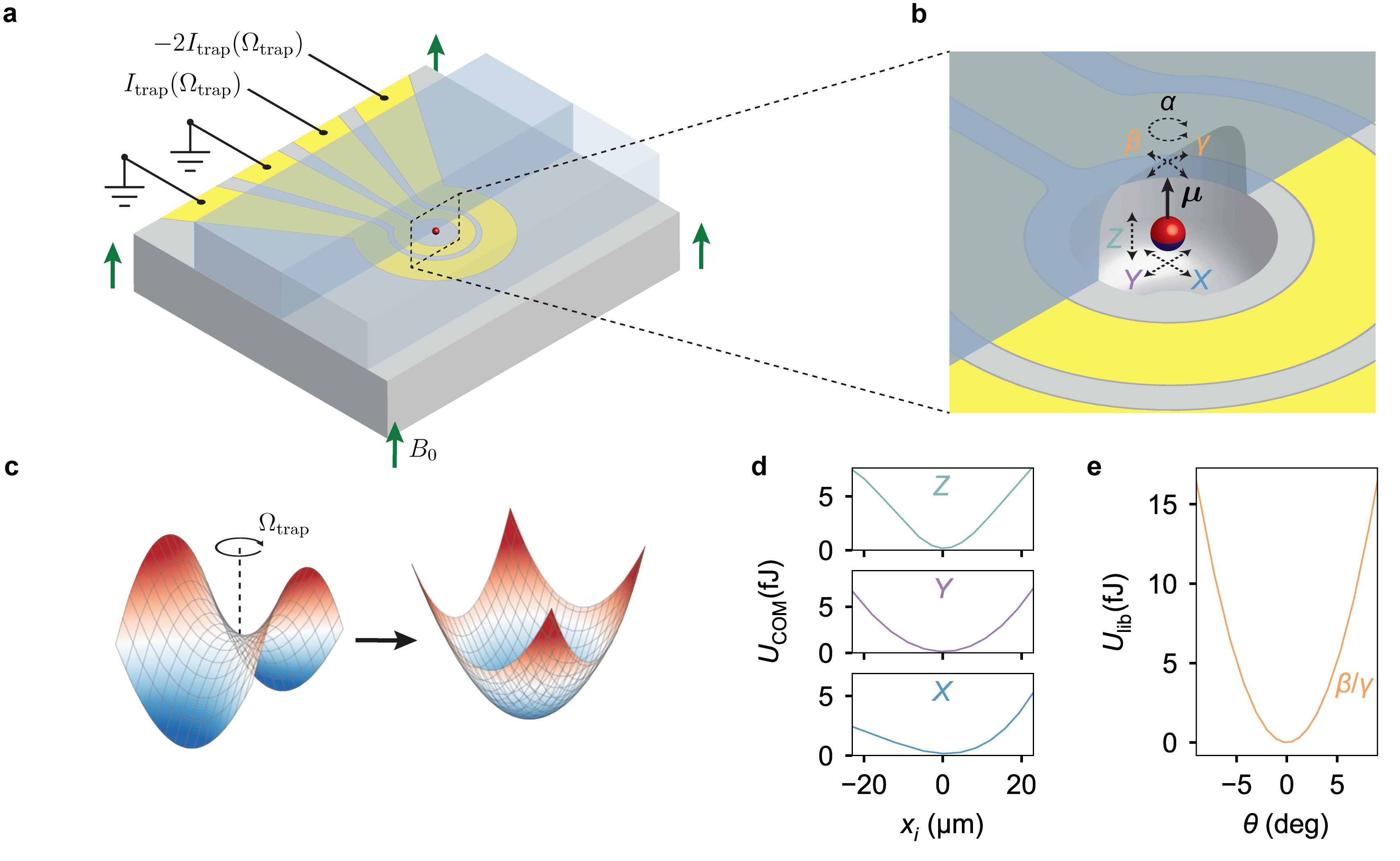}
    \caption{\textbf{Design and simulation of an on-chip magnetic Paul trap for ferromagnetic micro-particles.} \textbf{a}, A magnetic trapping field is generated by sinusoidal currents $I_\text{trap}$ at trapping frequency $\Omega_\text{trap}$, sent through two concentric conductive tracks (gold) on a chip (gray). A static magnetic field $B_0$ is applied in the z-direction, as defined by the coordinate frame shown, also defining various modes of rotation. \textbf{b}, A ferromagnetic particle with magnetic moment $\bm{\mu}$ is levitated in a pocket of the chip, with a (transparent) cover preventing escape during initial trapping. \textbf{c}, The rotating saddle potential caused by the alternating magnetic field results in a effective pseudo-potential $U_\text{COM}$ for the particle in the trap center, with the particle orientation fixed by $B_0$ and gravity compensated by a small gradient $\partial B_0/\partial z$. \textbf{d}, Finite element simulations of the geometry in \textbf{a}, showing $U_\text{COM}$ for motion along the x-, y and z-axis. \textbf{e}, Analytical simulation of the confining potential $U_\text{lib}$ in the rotational directions $\beta$ and $\gamma$, as a function of the angle $\theta$ between $B_0$ along the z-axis and the magnetic dipole moment $\bm{\mu}$.}
    \label{Figure1}
\end{figure*}

\section*{Magnetic Paul trap dynamics}\label{trap-eigenmodes}
A sinusoidal current $I_\text{trap}(\Omega_\text{trap})$ is applied to a planar current loop, with $\Omega_\text{trap}/2\pi$ the trap frequency (see Fig. \ref{Figure1}(a)). The current produces a magnetic field that interacts with the particle's permanent magnetic dipole moment $\bm{\mu}$. The particle orientation is fixed by a static field $B_0$ along the z-axis, produced by an external coil. Here the z-axis is the axis counter-aligned to gravity, see Fig. \ref{Figure1}(a-b). A small gradient $\partial B_0/\partial z$ compensates the gravitational pull. A current $-\xi I_\text{trap}(\Omega_\text{trap})$ sent through a second coplanar concentric current loop minimizes the magnitude of the alternating field along the vertical direction, improving stability and minimizing eddy current losses in the particle. Here $\xi$ is the current ratio (unless mentioned otherwise, $\xi = 2$).  The total alternating magnetic field is defined as $\bm{B_1}(\Omega_\text{trap})$. Averaging the magnetic energy of the particle over one trap field rotation results in an effective pseudo-potential \cite{perdriat_planar_2023}, see Fig. \ref{Figure1}(c):
\begin{equation}
    U_{\text{COM},i} = \frac{|\nabla_i (\mu_z B_{1,z})|^2}{4m\Omega_\text{trap}^2},
\end{equation}
with $i \in \{x,y,z\}$, $\mu_z$ the particle magnetic dipole moment along the z-axis, $B_{1,z}$ the amplitude of the time-dependent $\bm{B_1}$ field along the z-axis and $m$ the particle mass. Fig. \ref{Figure1}(d) shows the simulated pseudo-potential along the x-, y- and z-axis. Notice the asymmetry along the x-axis introduced by the loop slits, resulting in non-degenerate eigenfrequencies $\omega_x$ and $\omega_y$. In the secular regime (i.e. typically for stability parameter $q_i<0.4$), we can assume an effective static harmonic potential with translational eigenfrequencies
\begin{equation}
    \omega_i=\frac{q_i\Omega_\text{trap}}{2\sqrt{2}},
    \label{COM}
\end{equation}
where we define the stability parameter
    $q_i = 2B_{1,i}''B_{sat}/(\mu_0 \rho_m \Omega_\text{trap}^2),$
with $B_{1,i}'' = \partial^2B_{1,z}/\partial x_i^2$ and $x_i \in \{x, y, z\}$, $B_\text{sat}$ the particle remanent field, $\mu_0$ the magnetic permeability of vacuum and $\rho_m$ the particle density.\\

On top of the translational modes, the particle also has rotational degrees of freedom. We define rotations around the x-, y- and z-axes as $\gamma$-, $\beta$- and $\alpha$-rotations respectively. The static external field $B_0$ confines the $\beta$/$\gamma$-rotation resulting in particle libration with a potential energy $U_\text{lib} = \mu_z(\theta) B_0$, shown in Fig. \ref{Figure1}(e) as a function of the angle $\theta$ of the particle with the z-axis. The librational frequencies can be expressed as:
\begin{equation}
    \omega_\text{lib} = \sqrt{\frac{B_0B_{sat}V}{\mu_0I}},
    \label{lib}
\end{equation}
with $V$ the particle volume and $I$ the moment of inertia for the axis of rotation. For a spherically symmetric particle, $\alpha$-rotation is unconfined, and the $\omega_{\beta,\gamma}$-modes are degenerate. For angular stability $\omega_\text{lib}\gg \omega_i$ is required.\\

\begin{figure}[hbt!]
    \centering
    \includegraphics[page=1,scale=1]{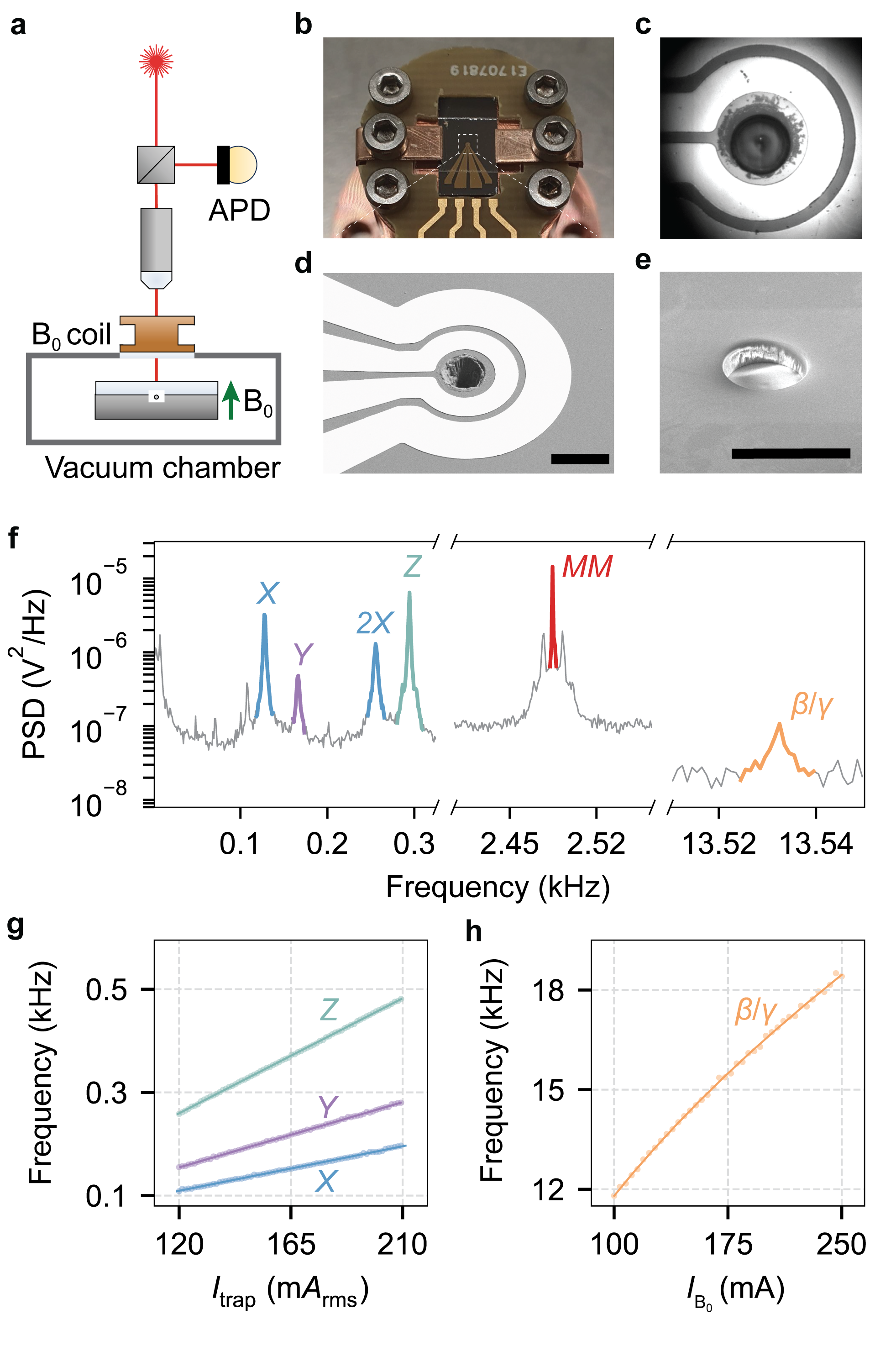}
    \caption{\textbf{Experimental implementation and tunable particle dynamics.} 
    \textbf{a}, The particle motion is measured in a vacuum chamber by collecting back-scattered laser light on an avalanche photodiode (APD) via an objective and 90/10 beam splitter. A coil on top of the chamber provides the static magnetic field $B_0$. 
    \textbf{b}, The chip is mounted on a printed circuit board for electrical interfacing and enclosed with a cover glass.
    \textbf{c} Video still of the particle levitating in the trap. Note that the particle levitates in the center of the current loops, which is slightly off-center in the pocket.
    \textbf{d}, Scanning micrograph of the chip with current tracks. The blind hole is produced by laser ablation. The scale bar is 100 $\mu$m.
    \textbf{e} Scanning micrograph of the cover glass with blind hole produced by focused ion-beam milling, providing a smoother surface for imaging. The scale bar is 100 $\mu$m.
    \textbf{f}, Power spectral density (PSD) of the APD signal without mechanical drive showing the translational and librational eigenmodes, the second harmonic $2\omega_x$ and the micromotion $\Omega_\text{trap}$. The right panel data was taken from a separate measurement to optimize the signal-to-noise ratio of the librational mode. 
    \textbf{g}, Translational eigenmodes $\omega_{x,y,z}$ as a function of trap current $I_\text{trap}$ with linear fit based on Eqn. \ref{COM}. 
    \textbf{h}, Librational eigenmode $\omega_\text{lib}$ as a function of external $B_0$ coil current $I_{\text{B}_0}$ with square-root fit based on Eqn. \ref{lib}. To increase signal-to-noise ratio, we drive the particle motion here.
    \vspace{15pt}
    }
    \label{Figure2}
\end{figure}

Finally, the particle exhibits micromotion at frequency $\Omega_\text{trap}/2\pi$. This intrinsic effect arises from the time-dependent confinement in the rotating magnetic field \cite{dehmelt_radiofrequency_1968}. This manifests itself in the frequency spectrum of the motion of the particle as mechanical modes spaced by multiples of the trap driving frequency $\Omega_\text{trap}$, with each of them presenting sidebands of the secular mechanical modes and their higher harmonics: $m\Omega_\text{trap} \pm n\omega_i$ ($n, m \geq 0$), with decreasing amplitude for increasing $n, m$. The micromotion spectral features can exhibit excess amplitude if the particle is displaced from the minimum of the pseudo-potential, for example as a result of gravity and stray magnetic fields (see SI, section II A).\\

\section*{Experimental implementation}
We implement the magnetic Paul trap on a silicon chip by patterning two gold loops of inner radii $60 ~\mu$m and $120 ~\mu$m and track widths of $50 ~\mu$m and $100 ~\mu$m, shown in Fig. \ref{Figure2}(c-d) (see SI, section I). A neodymium iron boron (NdFeB) microsphere of radius $a = 6.5$ $\mu$m is loaded in the blind hole with a tapered keratin probe to prevent scratching the soft gold tracks. A mechanically clamped cover glass on top of the chip provides enclosure of the trap without forming a vacuum-tight seal, enabling air removal. In the cover glass, a cylindrical hole of diameter $100$ $\mu$m and depth $15$ $\mu$m is milled with a focused-ion beam (see Fig. \ref{Figure2}(e)). The hole in the silicon and glass forms the pocket where the particle can levitate. After enclosing, the particle is magnetized by placing the sample in a $2.5$T magnetic field with which we expect to saturate the remanent field to at least 95 per cent of the maximum value \cite{magnequench_international_llc_technical_nodate}.\\

The chip is mounted on a custom-made printed circuit board for electronic connection as shown in Fig. \ref{Figure2}(b) and loaded in a vacuum chamber. We send a DC current $I_{\text{B}_0}$ through an external $B_0$ coil placed on top of the chamber, at a distance of 8.2 mm between the bottom of the $B_0$ coil and the center of the trap, to generate the static field $B_0$. We detect the motion of the particle optically, by illuminating the particle with a laser and detecting the back-scattered light with an avalanche photodiode. A schematic of the measurement setup is shown in Fig. \ref{Figure2}(a). The signal is filtered with a 100 kHz low-pass filter and measured with an oscilloscope. All scope traces were measured without driving the eigenmodes unless specifically mentioned otherwise. The laser power $P_\text{laser}$ incident on the particle is adjusted with an acousto-optic modulator between 20 nW and 1 $\mu$W.\\

As can be seen in Fig. \ref{Figure2}(f), the on-chip miniaturization of the trap allows for up to two orders of magnitude higher translational and librational eigenfrequencies in comparison to previous realizations due to increased magnetic field curvature $B_{1,i}''$ and decreased particle radius $a$, respectively \cite{sackett_magnetic_1993, perdriat_planar_2023, janse_characterization_2024}. In the spectrum, the three translational eigenmodes are observed in the APD signal at $P_\text{gas} = 1.0$ mbar, as well as the $2\omega_x$ harmonic. The translational eigenmodes are confirmed both by camera measurements and the phase response when driving. The slit along the x-axis in the trap geometry introduces a non-degeneracy of $\omega_y/\omega_x = 1.30$, which is in line with the expectation from simulations $\omega_y/\omega_x \propto B_{1,y}''/B_{1,x}'' = 1.32$ (see SI, section II B). In Fig. \ref{Figure2}(g) we see a linear scaling of the translational eigenfrequencies with the trap current $I_\text{trap} \propto B_1''$, as expected from Eqn. \ref{COM} (also see SI, section II B). The eigenfrequencies are extracted by fitting a Lorentzian lineshape to the power spectral density for each mode. The particle consistently escapes the trap for too high trapping current, i.e. $I_\text{trap} > 210$ m$\text{A}_\text{rms}$ for $\Omega_\text{trap}/2\pi$ = 2485 Hz. This is expected, since the stability parameter $q_i$ scales linearly with $B_{1,i}''$ and thus also linearly with $I_\text{trap}$, such that the trap enters the unstable regime $q_i \gtrsim 1$. Similarly, we find that the translational eigenfrequencies are inversely proportional to the trap frequency $\Omega_\text{trap}$ (see SI, section III).\\

The librational eigenmode is detected (see SI, section IV) with lower signal-to-noise ratio than the translational modes, which we attribute to detection being less sensitive to libration for a nearly spherical particle. The residual signal likely arises from surface roughness. Although two librational eigenmodes were found in a previous study for a cube magnet \cite{janse_characterization_2024}, here only one can be found, consistent with the expectations for a highly spherical particle. In Fig. \ref{Figure2}(h) we drive the particle motion to enhance the signal, and we recover a square-root dependence on the external coil current $I_{\text{B}_0}$, which is proportional to $B_0$, as expected (see SI, section V). Lastly, a confined $\omega_\alpha$-mode (rotation around the z-axis) is not identified, due to detection insensitivity, or particle and trap symmetry resulting in a lack of confinement, which is also found for electric Paul traps \cite{perdriat_rotational_2024}.\\

\begin{figure}[hbt]
    \centering
    \includegraphics[page=1,scale=1]{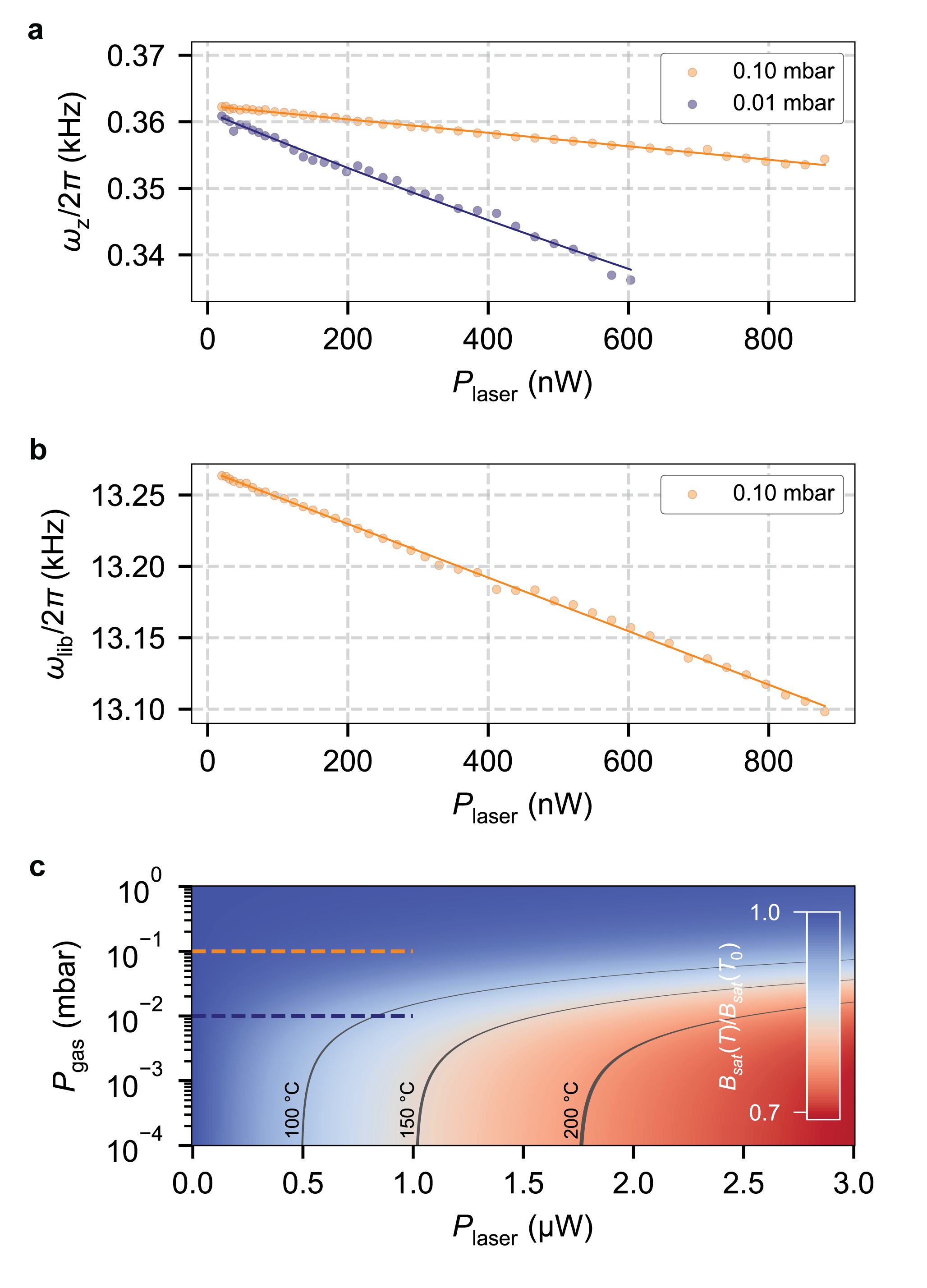}
    \caption{\textbf{Laser-induced particle thermodynamics.} 
    \textbf{a-b}, Translational eigenmode $\omega_z$ (\textbf{a}) and librational eigenmode $\omega_\text{lib}$ (\textbf{b}) as a function of laser power $P_\text{laser}$ irradiated on the particle for different values of $P_\text{gas}$. The eigenfrequencies are extracted by fitting a Lorentzian to the PSDs. The fit of the eigenfrequencies is based on the thermodynamical model described in SI, section VI with the absorptivity $\alpha$ and remanent field $B_\text{sat}(T_0)$ as fitting parameters. 
    \textbf{c}, Simulation of the change in remanent field $B_\text{sat}$ with respect to its initial value at room temperature $T_0$, as a function of gas pressure $P_\text{gas}$ and laser power $P_\text{laser}$. For the simulation the model in SI, section VI is used with the fit value of absorptivity $\alpha$ from the fit of $\omega_z$ to $P_\text{laser}$ at $P_\text{gas} = 0.10$ mbar. The orange and purple dashed lines indicate the $P_\text{laser}$ sweep range of panel \textbf{a-b} for both values of $P_\text{gas}$. The contour lines indicate the particle internal temperature.   
    }
    \label{Figure3}
\end{figure}

\section*{Vacuum operation}\label{thermodynics}
Interestingly, when operating the trap at lower gas pressures, we find that the particle dynamics are affected by the laser irradiation. This allows us to study the thermodynamical response of the ferromagnetic microparticle by varying the irradiated laser power $P_\text{laser}$ and the gas pressure $P_\text{gas}$. The laser power elevates the internal temperature $T$ of the ferromagnet. This changes the magnetization $M$ of the particle, where typically an increase in temperature results in a decrease of magnetization. For temperatures well below the Curie temperature, the change in magnetization is typically reversible and can be linearly approximated by $M(T) \approx M(T_0)\big[ 1 + \zeta_\text{th}(T-T_0) \big]$, with the thermal coefficient $\zeta_\text{th} = (1/M)dM/dT$ \cite{cullity_introduction_2009}. For our microparticles we expect $\zeta_\text{th} = -0.13 \%/\degree\text{C}$ \cite{magnequench_international_llc_technical_nodate}. Since $B_\text{sat}(T)=M(T)/\mu_0$, the temperature-induced decrease in magnetization leads to lower eigenfrequencies: $\omega_z \propto B_\text{sat}$ (Eqn. \ref{COM}) and $\omega_\text{lib} \propto \sqrt{B_\text{sat}}$ (Eqn. \ref{lib}). The  thermodynamical model for determining the particle internal temperature as a function of $P_\text{laser}$ and $P_\text{gas}$, which is used in Fig. \ref{Figure3}, is described in SI, section VI.\\

Fig. \ref{Figure3}(a-b) show how the translational eigenmode $\omega_z$ and the librational eigenmode $\omega_\text{lib}$ are affected by $P_\text{laser}$ at different values of $P_\text{gas}$. We can fit the data to our model, by leaving the remanent field $B_\text{sat}(T_0=300K)$ and effective absorptivity $\alpha$ of the particle as free parameters (see SI, section VI). We consistently find values for $B_\text{sat}(T_0)\approx 0.3~ \text{to} ~0.5$ T (using $\rho_m = 3.6\cdot10^3 ~\text{kg/m}^3$) and $\alpha\approx 0.1~ \text{to}~0.2$ for the various pressures, also in agreement with the values expected for our particle material (see SI, section V) \cite{magnequench_international_llc_technical_nodate, taylor_variation_1952, transmetra_gmbh_technical_nodate}. The SNR for $\omega_\text{lib}$ at $P_\text{gas} = 0.01$ mbar was too low to identify the mode consistently. Fig. \ref{Figure3}(c) shows how $B_\text{sat}$ is affected by $P_\text{laser}$ and $P_\text{gas}$, and allows us to identify the maximum laser power we can use without substantially affecting the particle dynamics.\\

\begin{figure}
    \centering
    \includegraphics[page=1,scale=1]{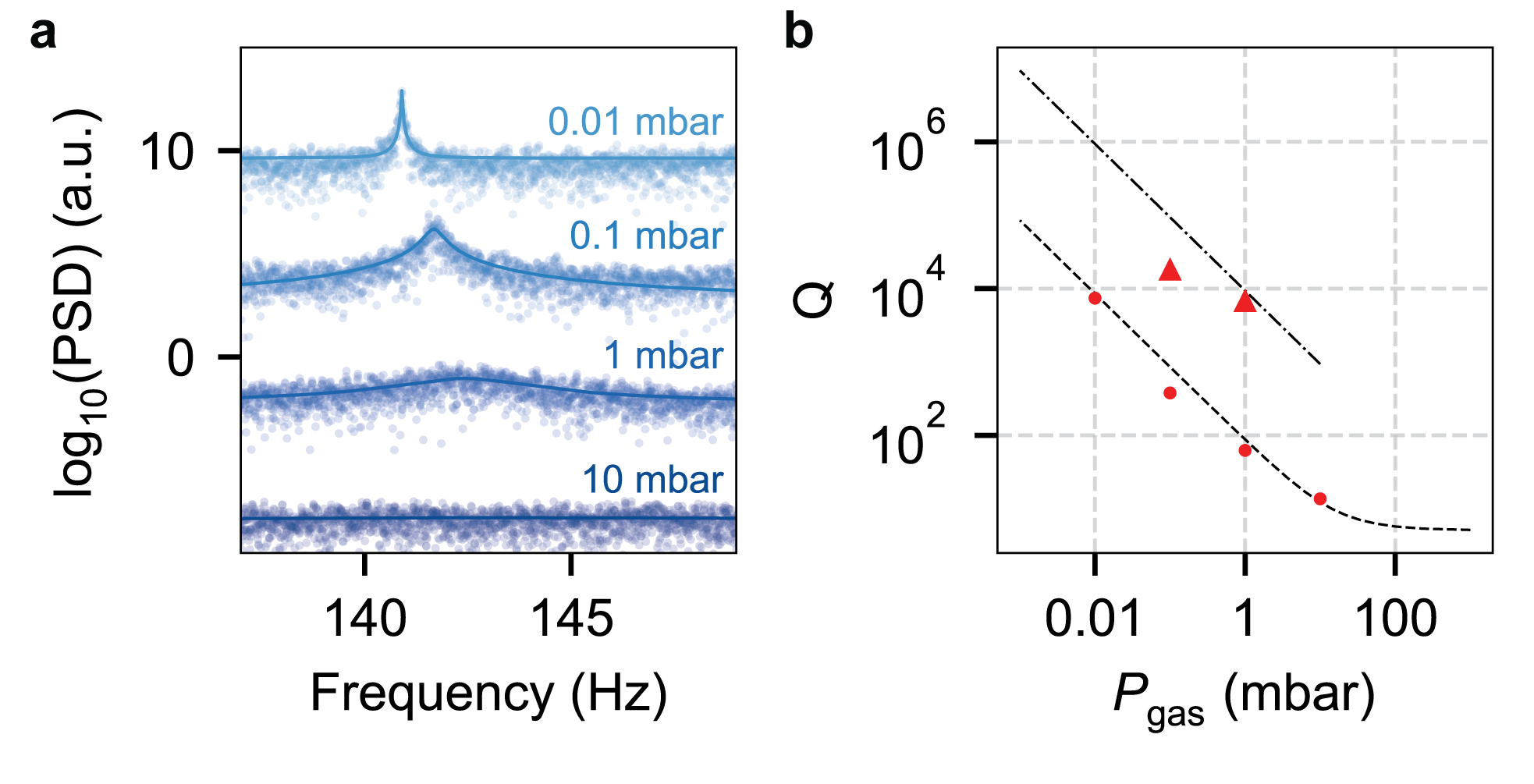}
    \caption{\textbf{Mechanical quality factor as function of gas pressure.} 
    \textbf{a}, Power spectral density of the $\omega_x$ mode for different values of gas pressure $P_\text{gas}$, vertically offset for clarity. Lines are a Lorentzian lineshape fit to the data from which we extract the linewidth. For decreasing gas pressure, the linewidth and eigenfrequency both decrease.
    \textbf{b}, Q-factor as a function of gas pressure $P_\text{gas}$ for the translational $\omega_x$ mode (circle) and librational $\omega_\text{lib}$ mode (triangle). All quality factors are determined from the Lorentzian linewidth. The dashed (dot-dashed) line is the expected Q-factor of the translational (librational) mode for a spherical particle limited only by gas damping (see SI, section VII A).The spectral densities are calculated from single time traces of 150 seconds.     
    }
    \label{Figure4}
\end{figure}

To quantify the isolation of the levitating particle from its environment, we measure the mechanical quality factors of the translational and librational eigenmodes. Fig. \ref{Figure4}(a) shows the Lorentzian lineshape fits of the $\omega_x$ eigenmode for different values of $P_\text{gas}$. As expected, we see a decreasing linewidth for lower gas pressure, see Fig. \ref{Figure4}(b). The translational eigenmodes follow the theoretical expectation \cite{millen_optomechanics_2020} down to at least $P_\text{gas} = 10^{-2}$ mbar (see SI, section VII A). Below this pressure our trap is affected by eigenfrequency fluctuations over timescales of seconds, resulting in non-Lorentzian lineshapes, which we attribute to instability in the trap current and laser power (see SI, section VIII). Moreover, since the particle is trapped in a blind hole enclosed by a glass cover, we expect the attainable gas pressure to be limited to $P_\text{gas}>10^{-3}$ mbar due to the combined effect of outgassing and confined geometry. Lower pressure could be attained by operating at cryogenic temperatures, or improving the evacuation geometry or materials. As validation we performed a ringdown measurement after multiple hours of turbo pumping, giving $Q \approx 4 \cdot 10^4$ for the $\omega_x$ mode (see SI, section IX). Since we cannot verify the value of $P_\text{gas}$, this quality factor might still follow the theoretical prediction of gas damping. However, other dissipation channels could also start to dominate for low enough $P_\text{gas}$ (see SI, section VII). The quality factor of the librational eigenmode could only be measured accurately between $P_\text{gas} =1-10^{-1}$ mbar, for which it is found to be consistent with gas damping. For higher and lower values of $P_\text{gas}$, we were unable to discern the signal of the undriven librational mode.\\ 

\section*{Proposal for spin-mechanical coupling}\label{spin-mech}
\begin{figure*}[!t]
		\centering
		\includegraphics[width=180mm]{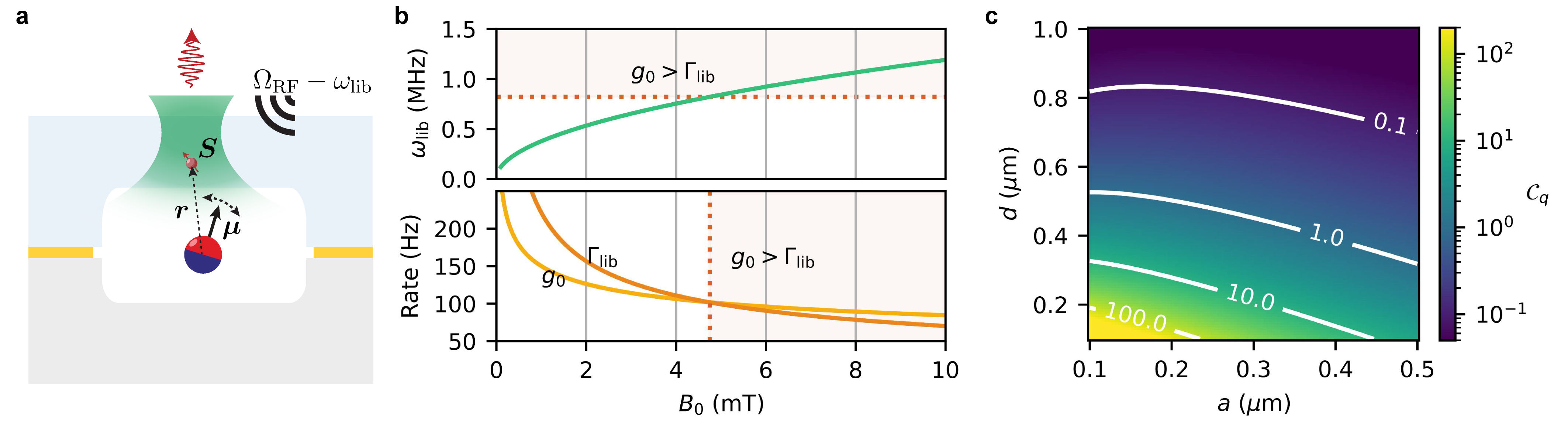}
		\caption{
        \textbf{Prospects for coupling the libration mechanical mode to a solid-state spin qubit.}
        \textbf{(a)} The top glass of the sample is replaced by diamond containing nitrogen-vacancy defect spins. The dipole-dipole interaction parametrically couples the particle libration and spin state. The spin is placed at a distance $ | \bm{r} |= a + d$ from the center of the ferromagnet. 
        \textbf{(b)} Influence of the external magnetic field $B_0$ on the librational eigenfrequency $\Ob$ (top panel), and the single spin coupling rate $g_0$ and mechanical heating rate $\Gb$ (bottom panel). Here we assume $a = 0.25 ~\mu\text{m}$, $d = 0.7~\mu\text{m}$, and operation of the setup at cryogenic temperature $T = 4~\text{K}$. Note how for large $\Ob$ the single spin coupling rate dominates over the mechanical heating rate. Combined with fast optical spin reset and a red-detuned microwave drive on the spin transition this allows cooling of the mechanical mode.
        \textbf{(c)} Quantum cooperativity $\Cq$ as a function of particle radius $a$, and (surface) distance $d$ between the magnet and the spin. For sufficiently small particles and distance to the spin we enter the single-spin strong coupling regime, allowing coherent exchange of spin-mechanical excitations.
        }
		\label{fig:sidebandCooling}
\end{figure*}

The stable orientation of the levitated particle's magnetic moment allows coupling to nearby spin-based quantum systems \cite{huillery_spin_2020}. This could enable spin-based readout of the particle motion (avoiding the use of a readout laser and associated particle heating), cooling of the particle motion, and ultimately preparation of non-Gaussian mechanical states. In the following paragraphs, we consider coupling of the librational eigenmode to the electronic spin of a single nitrogen-vacancy center in diamond, contained in the top cover of the trap, as depicted in Fig. \ref{fig:sidebandCooling}(a).\\

In Fig. \ref{fig:sidebandCooling}(b) we estimate the librational eigenfrequency $\Ob$ and single-spin coupling rate $g_0$ to a single spin oriented along the x-axis, placed at a distance $d = 0.7 ~\mu\text{m}$ from the surface of a particle with radius $a = 0.25\  \mu\text{m}$ (see SI, section X). For moderate external magnetic fields $B_0\approx5$ mT, the eigenfrequency exceeds the typical spin transition linewidth $\gs=2 \pi/T_2^*$ for isotopically purified diamonds ($T_2^*\approx 0.5$ ms \cite{maurer_room-temperature_2012, herbschleb_ultra-long_2019, abobeih_one-second_2018}) by orders of magnitude, while the coupling rate exceeds the expected mechanical heating rate $\Gb = \nth \gb$, considering a fixed linewidth $\gb = 2 \pi \cdot 1 ~\text{mHz}$ and $\nth = (\exp [\hbar \Ob/(\kB T_\text{bath})] - 1)^{-1}$ describing the thermal population of a bath at temperature $T_\text{bath} = 4 ~\text{K}$, with $\kB$ Boltzmann's constant. This puts the librational mode in the sideband-resolved regime of the spin transition ($\Ob > \gs$), and combined with the high coupling rate and fast optical spin initialization should allow cooling to the quantum regime.\\

Fig.~\ref{fig:sidebandCooling}(c) shows the quantum cooperativity $\Cq = 4 g_0^2/(\gs \Gb)$ as a function of magnet size $a$ and distance $d$ between spin and particle surface. Placing the spin at a surface distance $d < 0.4~\mu\text{m}$, and using a magnetic particle with radius $a = 0.25\  \mu\text{m}$ (corresponding to a picogram mass), would enable the system to work at the strong-coupling regime ($\Cq >1$). This would allow coherent transfer of quantum states between the spin and the mechanical mode, and the preparation of non-Gaussian mechanical states.\\

\section*{Conclusion}\label{conclusion}
We have demonstrated on-chip levitation of a ferromagnetic microsphere at room temperature, which enables robust confinement and control of relatively large (nanogram) masses, while giving access to high-frequency librational eigenmodes. We find that by reducing the readout laser power, the magnetization of the particle is maintained. We show that the quality factors of the translational eigenmodes are limited by gas damping down to at least $10^{-2}$ mbar. Assuming that libration is similarly affected by gas damping, quality factors for the librational eigenmodes on the order of $10^6$ are expected for pressure lower than $10^{-2}$ mbar. We project that the strong-coupling regime and sideband-cooling protocols are achievable with further miniaturization.\\  

The on-chip magnetic Paul trap could serve as a portable sensor for force/acceleration, pressure, and temperature. In addition, the thermodynamical model developed here, combined with pressure values derived from gas damping allows direct study of the thermal properties of magnetic materials in isolation.\\

Moreover, by combining smaller, non‑spherical particles with technical advances such as cryogenic operation and current stabilization, and by coupling to spin‑based quantum systems, we see a clear route towards quantum experiments with magnetically levitated picogram masses.

\section*{Data availability}
The data that support the findings of this study are available from the corresponding author upon reasonable request.

\section*{Acknowledgments}
We thank Maxime Perdriat for fruitful discussions regarding the detection scheme and Tjerk Oosterkamp for providing feedback on the manuscript. We also thank Evert Stolte for the design of the spin in Figure \ref{fig:sidebandCooling}.

\section*{Author contributions}
M.J. and B.H. conceptualized the experiment. M.J. designed the device. M.J., J.v.D. and R.W. fabricated the device. M.L.M. designed and realized the detection scheme. M.J., M.L.M and J.v.D. performed the measurements and analyzed the data. All the authors discussed the results. M.J. wrote the initial version of the manuscript and all authors contributed to writing the manuscript.

\section*{Funding}
This work was supported by the European Union (ERC StG, CLOSEtoQG, Project 101041115) and the Dutch Research Council (NWO, SUMMIT.1.016 Quantum Limits). M.L.M. wishes to thank the Swiss National Science Foundation for support (SNSF, Postdoc.Mobility Fellowship, grant nr. P500-2\_235435 / 1).\\

\section*{Competing interests}
The authors declare no competing interests.

\section*{Additional information}
\textbf{Supplementary information} Supplementary material is available online.\\

\textbf{Correspondence and requests for materials} should be addressed to Bas Hensen.\\

\section*{REFERENCES}
\bibliography{references.bib}

\end{document}


\title{On-chip levitation of ferromagnetic microparticles: \\Supplementary Information}

\date{April 2026}

\author{Martijn Janse\,\orcidlink{0000-0002-3877-9019}}
\author{M. Luisa Mattana\,\orcidlink{0009-0003-1793-0378}}
\author{Julian van Doorn\,\orcidlink{0009-0004-5892-8001}}
\author{Eli van der Bent\,\orcidlink{0009-0001-4270-8221}}
\author{Richard Wagner\,\orcidlink{0009-0009-8251-4462}}
\author{Robert Smit\,\orcidlink{0000-0003-2576-4851}}
\author{Bas Hensen\,\orcidlink{0000-0003-1754-6029}} \email{hensen@physics.leidenuniv.nl}
\address{Leiden Institute of Physics, Leiden University, P.O. Box 9504, 2300 RA Leiden, The Netherlands}

\maketitle

\section{Trap fabrication}\label{SM-trap-fab}
The on-chip trap is fabricated on a high-resistivity
silicon wafer, with the tracks patterned around a blind hole, with diameter and depth of 80 $\mu$m, produced via laser ablation (supplied by Potomac Photonics). We pattern the trap with electron-beam lithography and develop it, after which we sputter 5 nm of molybdenum-silicon (sticking layer) and 500 nm of gold. We finally lift off the excess gold in acetone. To facilitate visibility during FIB etching, we deposit a 5 nm thin film of electrically conductive material (platinum) on the cover glass, which we later remove at the hole during etching. 

\section{Simulation of magnetic fields ${B_0}$ and ${B_1}$}\label{SM-B0-B1-sims}
We have constructed a COMSOL Multiphysics$\textsuperscript{\textregistered}$ model of both the external coil to simulate $\bm{B_0}$ and the magnetic Paul trap to simulate $\bm{B_1}$ at the center of the magnetic Paul trap.

\subsection{Static magnetic field ${B_0}$}\label{SM-B0-sims}
\begin{figure}[h]
    \centering
    \includegraphics[page=1,scale=1]{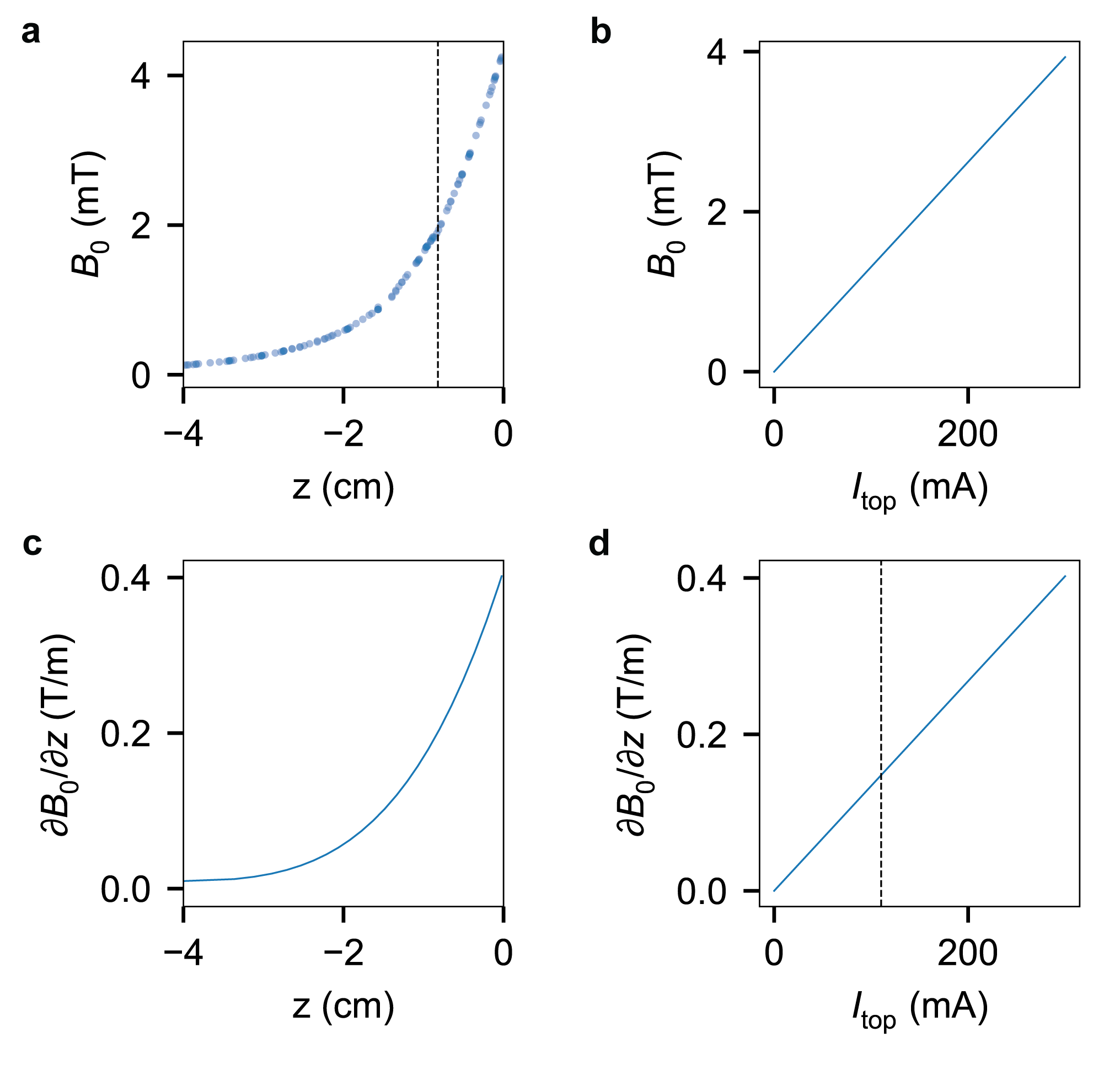}
    \caption{\textbf{Finite-element simulations of static magnetic field $B_0$ generated by the external coil}. \textbf{(a)} Magnetic field $B_0$ as a function of distance along the z-axis for $I_{\text{B}_0} = 150$ mA, where $z=0$ is set at the bottom of the external coil, resting on top of the vacuum chamber. The dashed line indicates the position of the sample at $z_0 = 8.2$ mm. 
    \textbf{(b)} Magnetic field $B_0$ as a function of current $I_{\text{B}_0}$ through the external coil at distance $z_0$, i.e. at the trap center.
    \textbf{(c)} Magnetic field gradient $\partial B_0/\partial z$ as function of distance along the z-axis for $I_{\text{B}_0} = 150$ mA.
    \textbf{(d)} Magnetic field gradient $\partial B_0/\partial z$ as function of current $I_{\text{B}_0}$ through the external coil at distance $z_0 = 8.2$ mm. The dashed vertical line marks the value of $I_{\text{B}_0}$ such that $\bm{F_G} = \bm{F_\text{DC}}$, where we use $B_\text{sat} = 0.30$ T (see \ref{SM-Bsat}). 
    }
    \label{FigureS2}
\end{figure}

The external coil has 835 windings of copper wire and is positioned at $x = y = 0$ mm and $z = z_0 = 8.2$ mm from the MPT center. At $I_{\text{B}_0} = 145$ mA we obtain $B_0 = 1.8$ mT (see Fig. S\ref{FigureS2}(a)) and $B_0' = \partial B_0 / \partial z = 0.19$ T/m (see Fig. S\ref{FigureS2}(c)). The values of $B_0$ and $B_0'$ scale linearly with $I_{\text{B}_0}$ (see Fig. S\ref{FigureS2}(b) and S\ref{FigureS2}(d), respectively).\\

On the one hand, the gravitational force pulling the particle downwards (in the negative z-direction) is given by $\bm{F_g} = m\bm{g}$ where $|\bm{g|} = 9.81$ m/$s^2$. On the other hand, the magnetic field gradient $\bm{B_0'}$ upwards (in the positive z-direction) is used to cancel the gravitional pull and acts with a force on the particle given by $\bm{F_\text{DC}} = \mu_z \bm{B_0'}$, where $\mu_z = B_\text{sat}V/\mu_0$. The dotted vertical line in Fig. S\ref{FigureS2}(d) shows the value of $I_{\text{B}_0}$ such that $\bm{F_g} = \bm{F_\text{DC}}$, where we have used the fitted value of $B_\text{sat} = 0.30$ T from the librational mode data (see \ref{SM-Bsat}).

\subsection{Alternating magnetic field ${B_1}$}\label{SM-B1-sims}
\begin{figure}[h]
    \centering
    \includegraphics[page=1,scale=1]{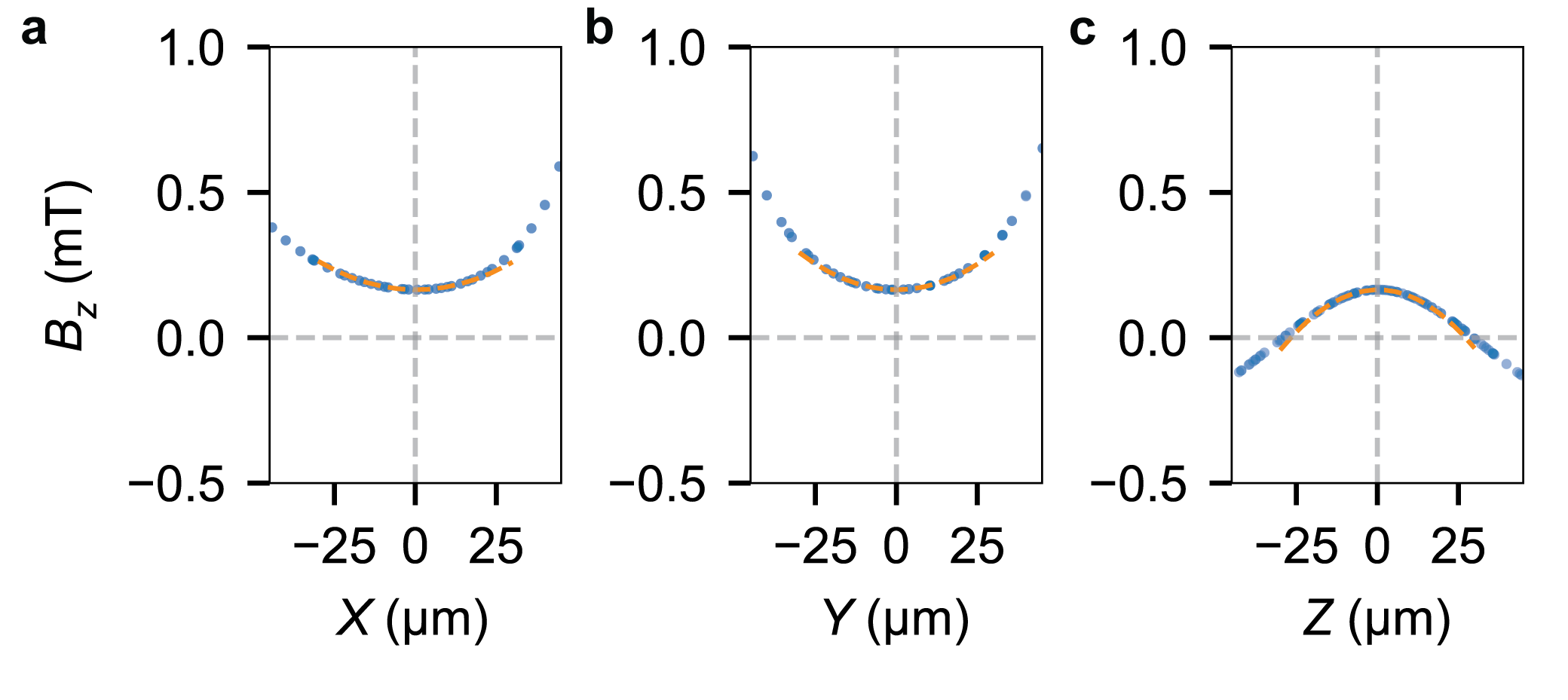}
    \caption{\textbf{Finite-element simulations of alternating magnetic field $B_1$ generated by the trap.} Amplitude of the z-component of the magnetic field $B_1$, i.e. $B_z$, along the \textbf{(a)} x-axis, \textbf{(b)} y-axis and \textbf{(c)} z-axis. The data points are finite-element simulations and the dashed line is a parabolic fit to extract the trapping stiffness.
    }
    \label{FigureS3}
\end{figure}

For simulating the alternating magnetic field $B_1$ we have constructed another COMSOL Multiphysics$\textsuperscript{\textregistered}$ model. The trap geometry is described in the main text. In the model we use a track conductivity of $4.56 \cdot 10^7$ S/m. The trap parameters are set to $I_\text{inner} = I_\text{outer}/2 = 163$ mA and $\Omega_\text{trap}/2\pi = 2485$ Hz.  Here we only take the z-component of the field, since the magnetic dipole moment of the particle is along the z-axis. The other relevant quantity is the magnetic field curvature $B_1'' = \partial^2B_{1,z}/dx_i^2$ with $x_i \in \{x, y, z\}$ from which we can derive the COM eigenfrequencies. Fig. S\ref{FigureS3} shows the simulated $B_1$-field within the trap. A harmonic potential is fitted to the center of these fields (between $x_i = -15 ~\mu$m and $15 ~\mu$m), where we assume a parabolic dependency on position. For the given trap parameters, this results in $B_1'' = \{2.15 \cdot 10^5, ~2.83 \cdot 10^5, ~4.54\cdot 10^5\}$ T/$\text{m}^2$. Both $B_1$ and $B_1''$ scale linearly with $I_\text{trap}$.\\

\section{Dependence of translational eigenmode on $\Omega_\text{trap}$}\label{SM-omega-trap-dependence}
\begin{figure}
    \centering
    \includegraphics[page=1,scale=1]{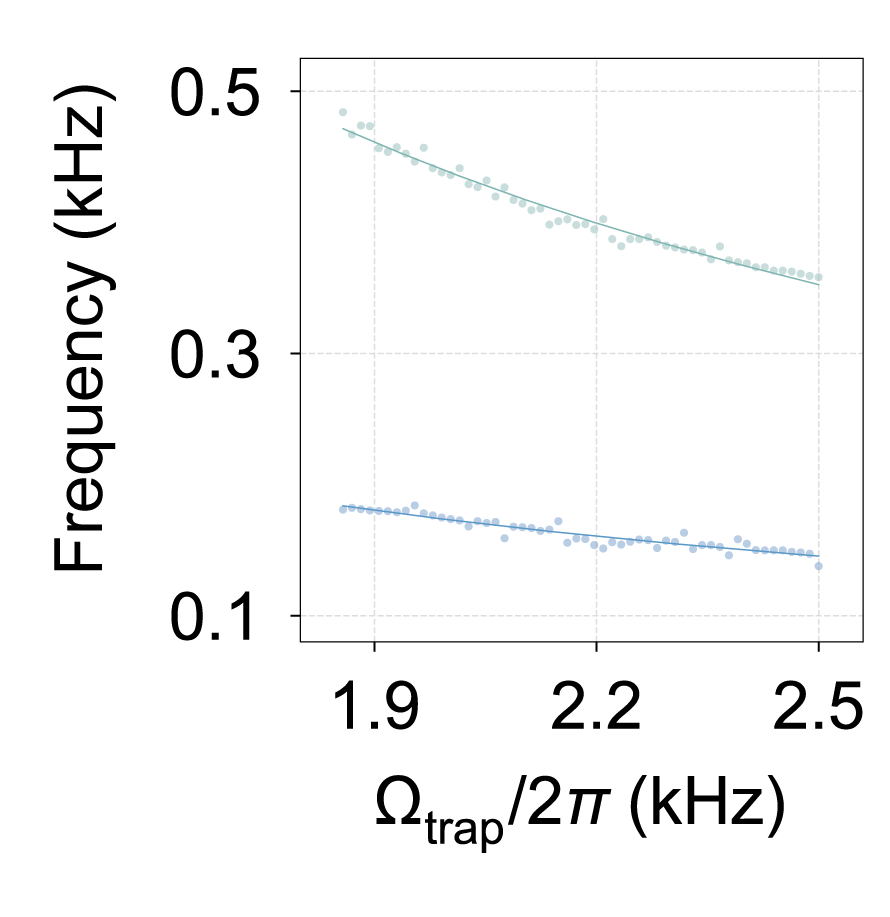}
    \caption{\textbf{Tunability of translational eigenmodes.} Dependence of translational eigenfrequencies $\omega_x$ and $\omega_z$ on trap frequency $\Omega_\text{trap}/2\pi$, in line with the expected $1/\Omega_\text{trap}$ relationship according to Eqn. 2 in the main text. The $\omega_y$ mode had too low SNR for fitting and is not shown.}
    \label{FigureS5}
\end{figure}

Fig. 2(g) in the main text shows the dependence of the translational eigenmodes on the trap current $I_\text{trap}$. However, according to Eqn. 2 in the main text, the translational eigenmodes can also be tuned by changing the trap frequency $\Omega_\text{trap}$, which is shown in Fig. S\ref{FigureS5} for the $\omega_x$- and $\omega_z$-eigenmodes. We retrieve the expected $1/\Omega_\text{trap}$ scaling. Due to low SNR, the peak frequencies were extracted from the PSD maxima instead of fitting a Lorentzian lineshape. As an artifact of the electrical circuit of the trap, the trap current was dependent on the trap frequency. Although we accounted for this by correcting the trap current when changing the trap frequency, this might still have introduced deviations from the $1/\Omega_\text{trap}$ scaling.

\section{Identification of librational eigenmode}\label{SM-lib-identification}
The eigenfrequencies of the librational eigenmode $\omega_\text{lib}$ are extracted by taking the maximum amplitude of the PSD, because not all driven peaks could be fitted with a Lorentzian lineshape (see Fig. 2(h) in the main text). Since the fundamental librational eigenfrequencies are in the kHz range, we were unable to verify these modes with camera measurements. Moreover, as an intrinsic feature of the (magnetic) Paul trap, we find spectral features at $\omega_\text{lib}\pm m \Omega_\text{trap}$.\\

Therefore, we confirm the fundamental librational eigenmode from the observed frequency shift $\Delta \omega_\text{lib}$ when sweeping the external coil current from $I_{\text{B}_0,i}$ to $I_{\text{B}_0,f}$. From Eqn. 3 in the main text we know that $\omega_\text{lib}(I_{\text{B}_0,i}) = C\sqrt{I_{\text{B}_0,i}}$ with $C = \Delta\omega_\text{lib}/\big(\sqrt{I_{\text{B}_0,f}}-\sqrt{I_{\text{B}_0,i}}\big)$. Filling in $\Delta\omega_\text{lib}/2\pi = 6.66$ kHz, $I_{\text{B}_0,i} = 100$ mA and $I_{\text{B}_0,f} = 250$ mA, we obtain as the fundamental frequency $\omega_\text{lib}(I_{\text{B}_0,i})/2\pi = 11.45$ kHz, close to the measured value at $11.81$ kHz. We again confirm this to be the fundamental librational eigenmode by performing an additional undriven scope measurement, sweeping the current from 150 mA to 160 mA. With this, we exclude other spectral features, e.g. those at $\omega_\text{lib}(I_{\text{B}_0,i})\pm m \Omega_\text{trap}$, from being the fundamental librational eigenmode.\\

\section{Determining the particle remanent field $B_\text{sat}$}\label{SM-Bsat}
The translational and librational eigenfrequencies both depend on the ratio of the particle remanent field $B_\text{sat}$ and the mass density $\rho_m$. Therefore, we cannot independently determine the value of $B_\text{sat}$ unless $\rho_m$ is known. In general, NdFeB has a mass density of $\rho_m = 7.4\cdot10^3 ~\text{kg/m}^3$ with a remanent field of $B_\text{sat} = 1.4$T, but the datasheet of the supplier reports a density of $3.6-4.2\cdot10^3 ~\text{kg/m}^3$ with a remanent field of $B_\text{sat} = 0.73-0.76$T \cite{magnequench_international_llc_technical_nodate}. However, the ratio $B_\text{sat}/\rho_m$ is the same in both cases.\\   

The value of $B_\text{sat}$ was determined from fitting the thermodynamical model (see \ref{SM-thermodynamics}) to the data in Fig. 3 in the main text. For the translational mode $\omega_z$ we find $B_\text{sat} = 0.50$ T both at $1.0$ mbar and $0.1$ mbar of gas pressure. For the librational mode $\omega_\text{lib}$ we find $B_\text{sat} = 0.29$ T at $1.0$ mbar. We attribute this discrepancy to the fact that, in order to fit $B_\text{sat}$, we need to use simulated values for the magnetic field curvature $B_1''$ (for the translational mode, see \ref{SM-B1-sims}) and for the static magnetic field $B_0$ (for the librational mode, see \ref{SM-B0-sims}), which both have their respective measurement and simulation errors, e.g. the error on the distance between sample and  external coil. Moreover, for the librational mode we assume the moment of inertia of a solid sphere, not accounting for a possible non-homogeneous mass distribution inside the particle.\\

We verify the value of $B_\text{sat}$ by rewriting Eqn. 3 as $\omega_\text{lib} = K\sqrt{B_\text{sat}}\sqrt{B_0}$ with $K = \sqrt{5/(2\mu_0\rho_ma^2)}$. We know that $B_0$ scales linearly with $I_{\text{B}_0}$, so $\Delta\omega_\text{lib} = K\sqrt{B_\text{sat}}(\sqrt{I_{\text{B}_0,f}/I_{\text{B}_0,i}\cdot B_{0,i}} - \sqrt{B_{0,i}})$. From this we retrieve $B_\text{sat} = 0.30$ T (assuming $\rho_m = 3.6\cdot10^3 ~\text{kg/m}^3$), which is in line with the value of $B_\text{sat}$ found from the fit of the data for the librational mode to the thermodynamical model, but lower than the value specified by the supplier (0.73-0.76 T).\\

Combining the COM eigenmode frequencies from Fig. 2(g) in the main text and the simulated $B_1''$ values, we can also calculate $B_\text{sat}$ from Eqn. 2 to verify the value with the data of the translational mode. For any value of $\omega_y$ and its corresponding value of $B_1''$ at $I_\text{trap}$, we find $B_\text{sat} = 0.45$T (assuming $\rho_m = 3.6\cdot10^3~\text{kg/m}^3$). For $\omega_x$ we find $B_\text{sat} = 0.45$ T and for $\omega_z$ we have $B_\text{sat} = 0.51$ T, which is in line with the value of $B_\text{sat}$ found from the fit of the data for the translational mode to the thermodynamical model, but again lower than the value specified by the supplier. The small discrepancy between the values derived from $\omega_{x,y}$ and $\omega_z$ can be explained by the fact that the ratio of the simulated magnetic field curvatures in the x-, y- and z-direction (1 : 1.32 : 2.11, whereas in the symmetric case, one would have 1 : 1 : 2) deviates from the measured difference in eigenfrequencies (1 : 1.33 : 2.5).\\

\section{Thermodynamics of a levitating ferromagnet}\label{SM-thermodynamics}
The COM and librational eigenfrequencies of the levitating ferromagnet scale with the remanent magnetic field, $B_\text{sat}$, as $\omega_i \propto B_\text{sat}$ and $\omega_\text{lib} \propto \sqrt{B_\text{sat}}$, respectively. However, the remanent field itself is temperature-dependent:

\begin{equation}
    B_\text{sat}(T) = B_\text{sat}(T_0)\big[1+\zeta_\text{th}(T - T_0) \big],
    \label{Bsat-thermo}
\end{equation}
where $T$ is the internal temperature of the ferromagnet, $T_0$ is room temperature and $\zeta_\text{th}$ is the temperature coefficient of $B_\text{sat}$ for ferromagnets. For our NdFeB magnet, we have $\zeta_\text{th} = -0.13\%/\degree\text{C}$, i.e. the magnet loses magnetization when heated. The temperature of the particle can be determined from a thermodynamical steady-state energy balance given by:

\begin{equation}
    P_\text{abs} = P_\text{rad} + P_\text{cond},
    \label{steady-state}
\end{equation}
where $P_\text{abs}$ is the laser power absorbed by the particle, $P_\text{rad}$ is the emitted power due to blackbody radiation and $P_\text{cond}$ is the power dissipated to the gas particles surrounding the particle.\\

The absorbed laser power $P_\text{abs}$ depends on the magnitude of $P_\text{laser}$, and on the absorptivity $\alpha$ of the particle at the wavelength of the laser $\lambda = 633$ nm. Since the particle is smaller than the spot size of the laser, we include a factor $\eta_g = 1-\exp(-2a^2/w^2)$, where we assume a Gaussian beam profile of waist $w$ of which we integrate the power over the particle radius $a$, assuming the particle to be centered in the profile:

\begin{equation}
    P_\text{abs} = \alpha(\lambda_\text{laser}) \big[1-\exp(-2a^2/w^2)\big]P_\text{laser}.
\end{equation}

For the radiated power $P_\text{rad}$ we assume the particle to be a blackbody radiator, such that we can describe the emission via the Stefan-Boltzmann law:

\begin{equation}
    P_\text{rad} = \epsilon(\lambda_\text{th})\sigma A \big( T^4 - T_\text{bath}^4 \big),
\end{equation}
where $\epsilon(\lambda_\text{th})$ is the emissivity of the particle at the thermal wavelength $\lambda_\text{th} \neq \lambda_\text{laser}$ set by Wien's law, $\sigma$ is the Stefan-Boltzmann constant, $A = 4\pi a^2$ the surface area of the particle and $T_\text{bath}$ the temperature of the bath, i.e. the gas surrounding the particle. \\

Lastly, for the power conducted to the gas, $P_\text{cond}$, we assume a dilute gas. In this free-molecular regime, the gas particles move ballistically and the Knudsen number $\mathrm{Kn} = l/L \gg 1$ with $l$ the mean free path and $L = a$ the characteristic length scale of the system. The conduction power is then given by \cite{hartnett_advances_1971}:

\begin{equation}
    P_\text{cond} = 3c_\text{acc}  A P_\text{gas} \sqrt{\frac{k_B}{2\pi m_g T_\text{bath}}}(T-T_\text{bath}),
\end{equation}
with $c_\text{acc}$ the thermal accommodation factor, $P_{gas}$ the gas pressure, $k_B$ the Boltzmann constant and $m_g$ the average molecular mass of air. The extra factor 3 follows from the assumption of a diatomic gas, approximately true for air, which has three translational and two librational degrees of freedom. The thermal accommodation factor is based on empirical data for air-metal surface interactions \cite{hartnett_advances_1971}.\\

Using the steady-state energy balance in Eqn. \ref{steady-state} we can solve for the internal temperature of the particle $T$. This temperature is then used to calculate $B_\text{sat}(T)$ via Eqn. \ref{Bsat-thermo}, which alters the translational (librational) eigenfrequency according to Eqn. 2 (Eqn. 3) in the main text. From the fits of this thermodynamical model to the experimental data shown in Fig. 3(a) in the main text, we find that $\alpha = 0.184 \pm 0.002$ (for $P_\text{gas} = 0.1$ mbar) and $\alpha = 0.116 \pm 0.004$ (for $P_\text{gas} = 0.01$ mbar). The fit on the data of the librational eigenmode $\omega_\text{lib}$ as a function of $P_\text{laser}$ in Fig. 3(b) in the main text returns $\alpha = 0.185 \pm 0.001$ (for $P_\text{gas} = 0.1$ mbar). The SNR for $\omega_\text{lib}$ at $P_\text{gas} = 0.01$ mbar was too low to identify the mode.\\ 

The fit value of $\alpha$ is an effective absorptivity, which does not include the reflections and diffusion caused by the vacuum chamber window and trap cover glass. In practice, less laser light reaches the particle, making the fit value of $\alpha$ a lower bound. The value of $\alpha \in [0,1]$ is dependent on wavelength and material, as well as on surface roughness and oxidization, making it hard to compare to literature values. However, typical values found for iron are 0.2-0.8 \cite{taylor_variation_1952, transmetra_gmbh_technical_nodate}. This makes the effective absorptivity values from the fits reasonable.\\

The discrepancy in fit values of the absorptivity $\alpha$ might be explained by the fact that the gas pressure is measured with a gauge before the vacuum chamber and, additionally, that the top cover glass might introduce pressure gradients between the position of the particle and the rest of the vacuum chamber. This could result in a discrepancy between the measured $P_\text{gas}$ and the actual $P_\text{gas}$ at the particle.\\

There are some limitations to using this model to fit the scaling of the COM and librational eigenfrequencies with $P_\text{laser}$ and $P_\text{gas}$. Firstly, the absorptivity $\alpha$ and emissivity $\epsilon$ of our NdFeB particle are unknown. The values depend on the material as well as, for example, on surface roughness and oxidization. Therefore, we treat $\alpha$ and $\epsilon$ as fit parameters. In the fit, we impose the simplifying assumption that $\alpha(\lambda_\text{laser}) = \epsilon(\lambda_\text{th})$. Strictly speaking, Kirchhoff's law of thermal radiation only justifies this equality at identical wavelengths, so this condition is not physically rigorous in our case. However, treating $\alpha$ and $\epsilon$ as independent fit parameters leads to strong correlation between $\alpha$ and $\epsilon$ that prevent a meaningful fit, making this assumption necessary in practice. Namely, in this model the eigenfrequencies depend on temperature, while both $\alpha$ (laser heating) and $\epsilon$ (radiative cooling $\propto T^4$) affect the temperature in opposite ways. Another limitation is uncertainty in the beam waist (measured with a camera) and the assumption of a Gaussian beam profile, since the beam shape might be deformed by interaction with the trap cover glass.

\section{Dissipation mechanisms}\label{SM-dissipation}
As discussed in the main text, gas damping seems to limit the Q-factors until pressures of at least $10^{-2}$ mbar. Here, we discuss several dissipation channels that could limit the Q-factor of the levitating particle at lower pressures.

\subsection{Gas damping}\label{SM-dissipation-gas}
For a spherical particle of radius $a$ and mass $m$ the gas damping rate for the translational degrees of freedom $\gamma_\text{COM}$ is given by \cite{millen_single_2018, millen_optomechanics_2020, beresnev_motion_1990}:

\begin{equation}
    \frac{\gamma_\text{COM}}{2\pi} = 3\mu_\nu (a/m)\frac{0.619}{0.619+\text{Kn}}(1+c_K),
\end{equation}
with $c_K = 0.31\text{Kn}/(0.785+1.152\text{Kn}+\text{Kn}^2)$, $\text{Kn} = l/a$ the Knudsen number with $l = k_B T_\text{bath}/(\sqrt{2}\sigma_\text{gas}P_\text{gas})$ the free mean path, $k_B$ the Boltzmann constant, $T_\text{bath}$ the gas bath temperature, $P_\text{gas}$ the gas pressure and $\sigma_\text{gas} = \pi d_m^2$ with $d_m = 0.372$ nm the (average) diameter of air molecules. $\mu_\nu$ is the viscosity coefficient for a dilute gas given by $\mu_\nu = 2\sqrt{m_\text{gas}k_BT_\text{bath}}/3\sqrt{\pi}\sigma_\text{gas}$ with $m_\text{gas}$ the (average) mass of air molecules.\\

If we assume low pressures such that $\text{Kn} \gg 1$, one can also obtain an analytical expression for the gas damping rate of a spherical particle for the librational degrees of freedom, $\gamma_\text{lib}$ \cite{millen_single_2018, millen_optomechanics_2020, martinetz_gas-induced_2018}:

\begin{equation}
    \frac{\gamma_\text{lib}}{2\pi} = \frac{30c_\text{acc} \mu_\nu \sigma_\text{gas}P_\text{gas}}{8\pi\sqrt{2}k_B T_\text{bath}\rho_m a},
    \label{Gamma_lib}
\end{equation}
where $c_\text{acc}$ is the thermal accommodation factor, which we approximate for air by taking $c_\text{acc} = 0.65$ for nitrogen gas \cite{millen_single_2018} and $\rho_m$ the mass density of the particle. In Fig. 4(b) in the main text we have only plotted \ref{Gamma_lib} for $\text{Kn} > 100$ to make sure we meet the requirement of low pressure.

\subsection{Eddy current damping in the loops}\label{SM-dissipation-eddy}
\begin{figure}
    \centering
    \includegraphics[page=1,scale=1]{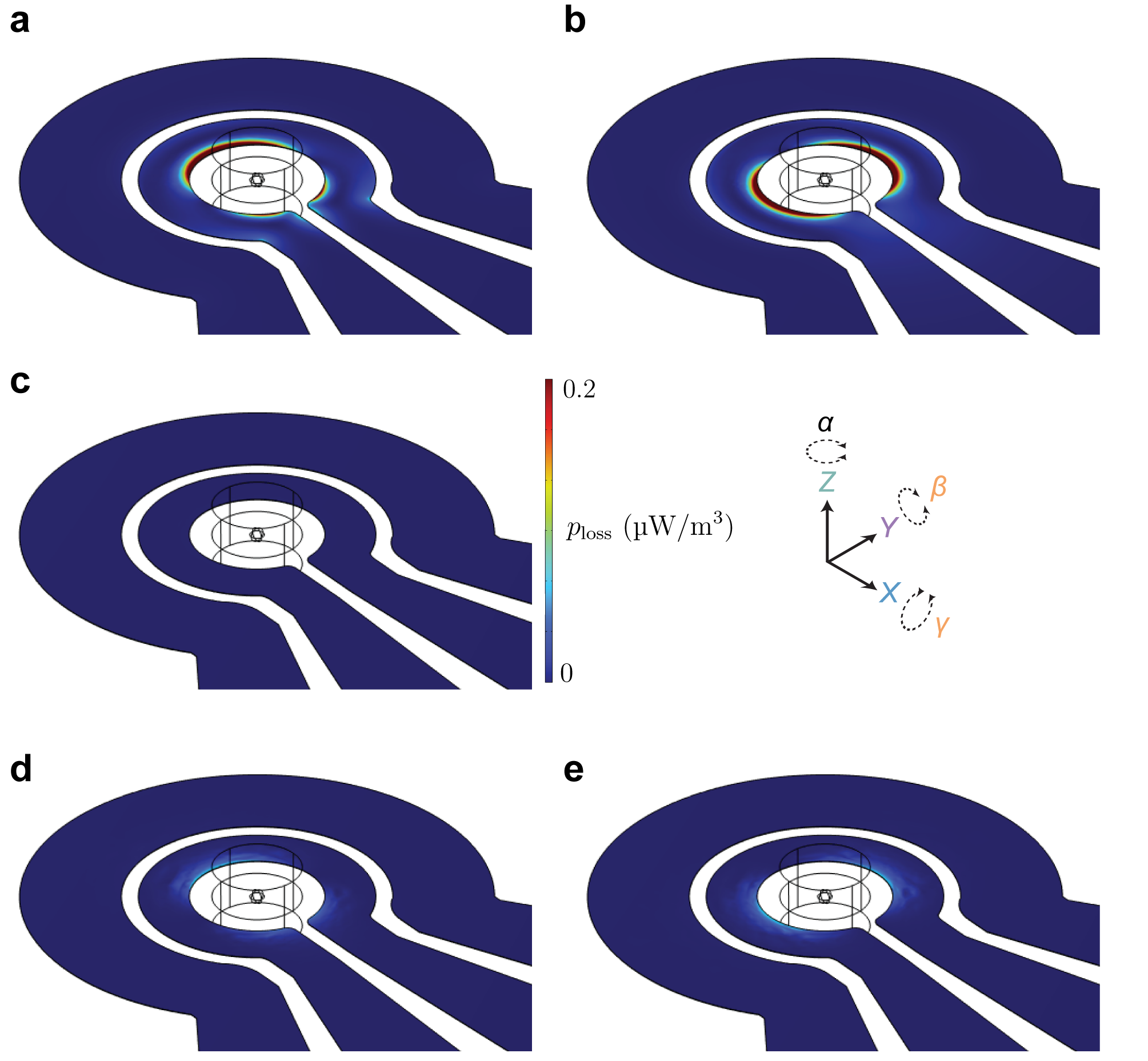}
    \caption{\textbf{Finite-element simulations of eddy current damping for translational and librational eigenmodes.} Simulation of the dissipated energy via induced eddy currents in the gold trap tracks (volumetric loss density $p_\text{loss}$) of the ferromagnetic particle oscillating along the \textbf{(a)} x-axis, \textbf{(b)} y-axis and \textbf{(c)} z-axis. The dissipated energy of the libration eigenmodes is also given in the \textbf{(d)} $\beta$-direction and \textbf{(e)} $\gamma$-direction. The kinetic energy of the particle as a result of the Lorentz (angular) Velocity is $E_\text{kin} = 2\cdot10^{-18}$ J in all panels.
    }
    \label{FigureS7}
\end{figure}

To estimate the dissipation due to eddy currents induced by the moving magnetic particle in the gold tracks of the trap, we model the trap geometry in COMSOL Multiphysics$\textsuperscript{\textregistered}$ software. Induced eddy currents are modeled by including a static magnetic particle with radius R = $6.5$ $\mu$m, density $\rho_m = 3.6\cdot10^3$ kg/$\text{m}^3$, remanent field $B_{\text{sat}}$ = $0.5$ T (see $\ref{SM-Bsat}$), and assigning a Lorentz (angular) Velocity term to the conducting track domains surrounding the particle, i.e. simulating relative motion of the tracks in the rest frame of the particle. Calculating the resulting electromagnetic losses $P_\text{loss}$ from the volumetric loss density $p_{\text{loss}}$ (see Fig. S\ref{FigureS7}), we find a quadratic dependence on the velocity in each direction (x, y and z), as expected.\\

We estimate the resulting quality factor $Q_i = 2\pi f_i \cdot (\frac{1}{2} m v_i^2)/P_{\text{loss},i}$, taking the mass $m$ of the particle and $f_i$ the frequencies of the eigenmodes along the Cartesian directions $i = \{x,y,z\}$. The tracks are assumed to have a conductivity of $4.56 \cdot 10^7$ S/m. For $I_\text{inner} = I_\text{outer}/2 = 160$ mA and $\Omega_\text{trap}/2\pi = 2485$ Hz we obtain $f_i = \{150, ~200, ~360\}$ Hz. Therefore, if the Q-factors would only be limited by eddy current damping, we would have $Q_i = \{3.2\cdot10^6, ~3.0 \cdot 10^6, ~7.9\cdot 10^9\}$. Similarly, for the librational eigenmodes $\omega_\beta$ and $\omega_\gamma$ (named $\omega_\text{lib} = \omega_\beta = \omega_\gamma$ in the main text due to the eigenfrequency degeneracy), we find $Q_\beta = 3.9\cdot10^9$ and $Q_\gamma = 3.4 \cdot 10^9$.\\

\subsection{Eddy current damping in the magnet}\label{sm-dissipation-eddy-magnet}
The magnetic field $\bm{B_1}$ of the trap also induces eddy current in the levitating ferromagnet. Although we minimize the magnetic field at the center of the trap by using two current loops, the particle still has a finite volume and is affected by the magnetic field (see \ref{SM-B1-sims}). The induced current density $j$ in the particle is given by \cite{perdriat_planar_2023}:

\begin{equation}
    j_\text{ind} \approx \Omega_\text{trap}\sigma_\text{el}a^3B_1'',
\end{equation}
where $\sigma_\text{el}$ is the electrical conductivity of the magnet. The dissipated power is then given by:

\begin{equation}
    P_\text{ind} \approx \Omega_\text{trap}^2 \sigma_\text{el}a^9 (B_1'')^2
\end{equation}
Using $\Omega_\text{trap} = 2485$ Hz, $\sigma_\text{el} = 0.67\cdot10^6$ S/m, $a = 6.5 ~\mu$m and $B_1'' = 2.15 \cdot 10^5$ T/$\text{m}^2$ (from the simulations in the x-direction, see \ref{SM-B1-sims}), we obtain $P_\text{ind} = 4\cdot10^{-24}$ W. For the $\omega_x$-, $\omega_y$- and $\omega_\text{lib}$-modes, this is three orders of magnitude lower than the loss due to the eddy current damping in the loops, so this dissipation channel would not limit the Q-factor further. For the $\omega_z$-mode, it is of comparable magnitude.

\section{Eigenfrequency fluctuations}\label{sm-flucts}
For gas pressures lower than $10^{-2}$ mbar, we observe eigenfrequency fluctuations on the order of several hertz over timescales of seconds. This results in non-Lorentzian lineshapes of the eigenmode, which prevents us from determining the quality factor from the spectrum. There are several potential causes of these eigenfrequency fluctuations. First, it could be due to instabilities in trap current, which is linearly proportional to $B_1''$ and thus directly affects the translational eigenfrequencies via Eqn. 2 in the main text. Second, fluctuations in the laser power could lead to fluctuations in particle internal temperature, altering $B_\text{sat}$, as described in \ref{SM-thermodynamics}. This effect is more pronounced for lower gas pressures.\\

\begin{figure}[h]
    \centering
    \includegraphics[page=1,scale=1]{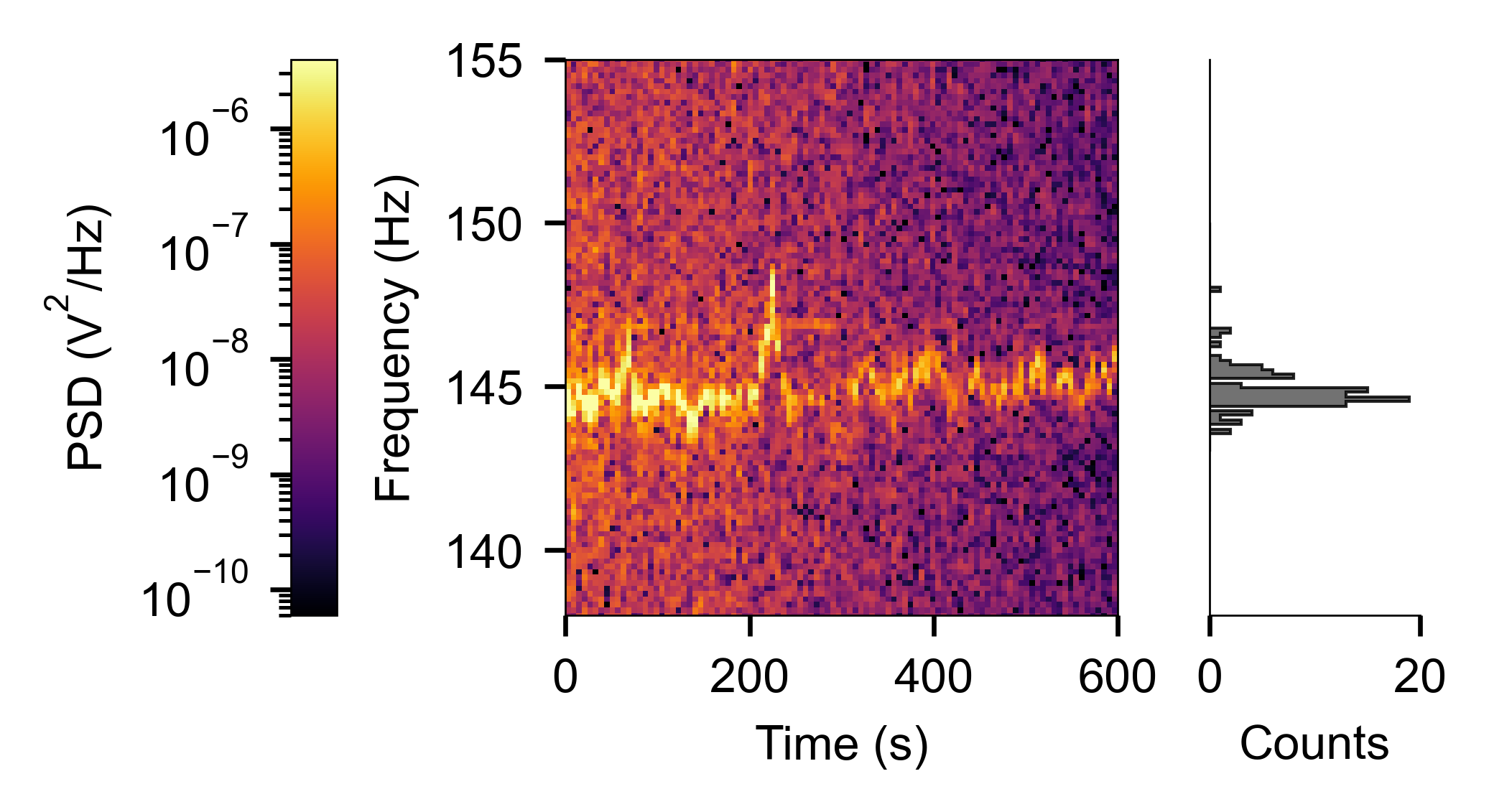}
    \caption{\textbf{Fluctuations of eigenmode frequency.} Spectrogram of the translational eigenmode $\omega_x$ as a function of time at $P_\text{gas} = 10^{-3}$ mbar and $P_\text{laser} = 200$ nW. The histogram shows the distribution of peak frequencies of every segment of the spectrogram.  
    }
    \label{FigureS6}
\end{figure}

In Fig. S\ref{FigureS6} we show the eigenfrequency fluctuations of the $\omega_x$-mode for a 600-second time trace measured at $P_\text{gas} = 10^{-3}$ mbar and at $P_\text{laser} = 200$ nW. The time trace is cut up in 100 segments of 6 seconds each, after which a PSD was calculated. Due to limited resolution in every segment, the peak frequency was estimated from the PSD maximum of every segment. The histogram shows the eigenfrequency fluctuations on the order of several hertz.\\

\section{Ringdown measurement}\label{sm-ringdown}
As a cross-check for the quality factors determined from the spectral linewidth (see Fig. 4 in the main text), we also perform a ringdown measurement on the translational eigenmode $\omega_x$. In order to minimize the gas pressure, we turbo pump the vacuum chamber for multiple hours before performing the ringdown. However, as mentioned in the main text, we cannot reliably determine the gas pressure in the pocket of the trap. We excite the translational eigenmode $\omega_x$ by sending a current pulse through an external coil (different than the $B_0$ coil on top of the vacuum chamber).\\

\begin{figure}
    \centering
    \includegraphics[page=1,scale=1]{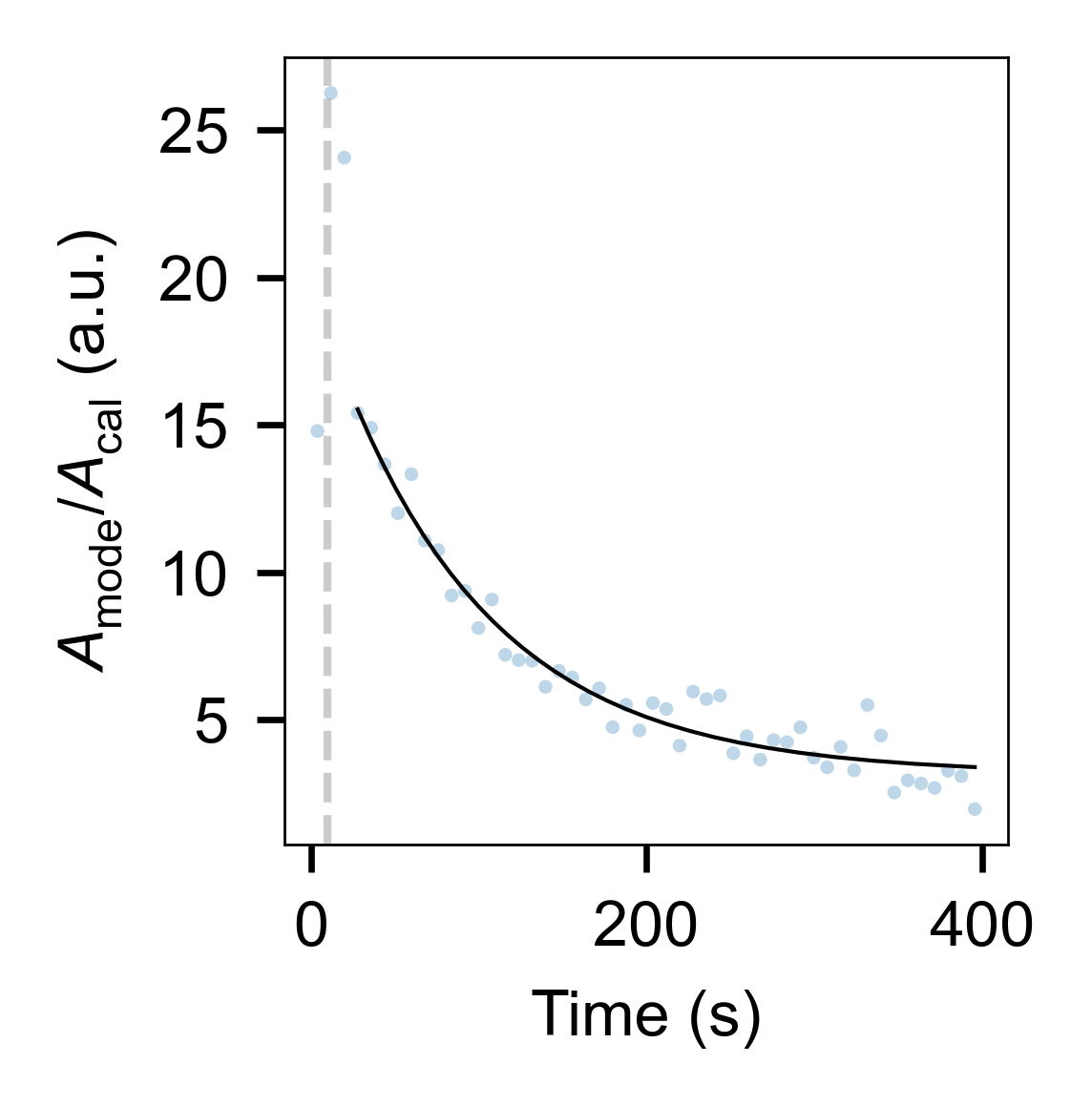}
    \caption{\textbf{Ringdown measurement of translational eigenmode.} Amplitude ratio $A_\text{mode}/A_\text{cal}$ as a function of time, where $A_\text{mode}$ is the amplitude of the $\omega_x$ mode and $A_\text{cal}$ the amplitude of a calibration tone which we send at fixed frequency to correct for changes in overall SNR. Both amplitudes are obtained from the PSD and the ratio is averaged over time in 50 bins. The black line is an exponential decay function fit to the data to extract the ringdown time $\tau$. The vertical dashed line indicates the moment when the current through the external coil to excite the particle is turned off. The two data points directly after the excitation are not included in the fit because of non-linear response at high amplitude.     
    }
    \label{FigureS4}
\end{figure}

The ringdown analysis is done by first calculating the PSD of a 400-second time trace of the APD signal. We integrate the PSD in a bandwidth of 3 Hz around the peak frequency and take the square root as a measure for the mode amplitude $A_\text{mode}$. This is an indirect measure, because the voltage PSD is not calibrated to a position PSD. We then correct for SNR fluctuations by dividing the amplitude by the square root of a PSD integral of a constant calibration tone sent by the external coil $A_\text{cal}$, in a bandwidth of 600 mHz. This ratio $A_\text{mode}/A_\text{cal}$ is then averaged in 50 bins of 8 seconds, of which the data points are shown in Fig. S\ref{FigureS4}. We fit to the data an exponential decay function $A_\text{mode}/A_\text{cal}(t) = A_\text{mode}/A_\text{cal}(t_0)\exp(-t/\tau) + C$ with $A$ the corrected amplitude, $t$ the time, $t_0$ the moment of the excitation, $\tau$ the ringdown time and $C$ a constant offset. From this, we obtain the fitting parameter $\tau$, from which we calculate $Q = \omega_x\tau/2$. We find that $Q \approx 4\cdot10^4$ after several hours of turbo pumping.\\

\section{Spin-mechanical coupling}\label{sm-spin-mech}
In the following, we discuss the derivation of the equation describing the coupling between the mechanical librational mode of the levitated magnet, and the spin of a nitrogen-vacancy (NV) defect. The single electronic spin in the NV defect center is placed at a distance $\bm{r}$ from its center of mass. We define $|\bm{r}| = a + d$, with $d$ the distance between the spin and the surface of the magnet. To simplify the calculation, we assume the spin to be oriented parallel to the x-axis. The total magnetic field at the location of the spin $\bm{r}$ is the result of the contribution from the (static) external field $\Bext$, and the field $\Bm$ generated by the magnet. The latter is a function of the orientation $\theta$ of the magnet, modulated in time with frequency $\Ob$ due to libration. This modulation in turn induces a variation of the spin transition frequency $\Omega_\text{RF}$. In the following derivation, we assume that the magnet is librating with a small amplitude around the y-axis, such that (applying the small-angle approximation), the permanent magnetic dipole vector $\bm{\mu}$ is oscillating with amplitude $\theta$ along the x-axis.\\

In analogy to cavity optomechanical systems, in which the motion of a resonator induces a change in the cavity frequency, we model the relation between the spin transition frequency and the libration motion through a spin-mechanical coupling mediated by the dipole-dipole interaction. We calculate the single-spin coupling rate $g_0$ to be \cite{huillery_spin_2020}:

\begin{equation}
    g_0 = \frac{1}{6} \gamma_e \mu_0  M \frac{a^3}{(a + d)^3} \betazpf,
\end{equation}
where $\gamma_e/(2 \pi) = 28~\text{GHz}$ is the gyromagnetic ratio of the spin electron, and $\betazpf = \sqrt{\hbar/(2 I \Ob)}$ is the zero-point fluctuation of the libration mode. Additionally, as in the main text, we define here the quantum cooperativity $\Cq = 4 g_0^2/(\gs \Gb)$ using the typical spin transition linewidth $\gs=2 \pi/T_2^*$, and the mechanical heating rate $\Gb = \nth \gb$, considering a fixed linewidth $\gb = 2 \pi \cdot 1 ~\text{mHz}$ and $\nth = (\exp [\hbar \Ob/(\kB T_\text{bath})] - 1)^{-1}$ the thermal population of a bath at a temperature of $4 ~\text{K}$, where $\kB$ is Boltzmann's constant.\\

In our proposed scheme, a 532 nm laser continuously polarizes the spin in the ground state $\ket{m_s= 0}$, bringing the system in the total state $\ket{m_s= 0} \ket{n}$, where $\ket{n}$ is the phonon population of the librational mode. We then apply a microwave tone, red detuned from the transition $\ket{m_s = 0} \rightarrow \ket{m_s = - 1}$ by an amount $\Ob$. This is equivalent to selecting the anti-Stoke process, bringing the total state of the system form $\ket{m_s = 0}\ket{n}$ to $\ket{m_s = -1}\ket{n - 1}$, decreasing the mechanical mode phonon population. Provided that the cooling rate $\gamma_\text{cool} = C_q/(\gb \nth)$ and the spin re-initialization rate are both faster than the mechanical decoherence rate, repeating the protocol will eventually lower the phonon population to its quantum ground state $\ket{n = 0}$.\\ 

\section*{REFERENCES}
\bibliography{references.bib}